\documentclass[onecolumn, 12pt, journal]{IEEEtran}
\ifCLASSINFOpdf
\else
\fi

\usepackage{graphicx}
\usepackage{subfig}
\usepackage{amsmath}
\usepackage{algorithm}
\usepackage{algorithmic}
\usepackage{cite}
\usepackage{amsthm}
\usepackage{amssymb}
\usepackage{color}
\usepackage{multicol}
\usepackage{bm}
\usepackage{tabularx}
\usepackage{booktabs}

\theoremstyle{definition}

\theoremstyle{remark}

\theoremstyle{proposition}

\begin{document}
%
\title{Information-Centric Wireless Networks with Virtualization and D2D Communications}

\author{\IEEEauthorblockN{Kan Wang\IEEEauthorrefmark{1}, F.~Richard~Yu\IEEEauthorrefmark{2}, Hongyan Li\IEEEauthorrefmark{1}, and Zhengquan Li\IEEEauthorrefmark{3}\IEEEauthorrefmark{4}}\\
\IEEEauthorblockA{\IEEEauthorrefmark{1}State Key Lab. of Integrated Services Networks, Xidian University, Xi'an, P.~R.~China}\\
\IEEEauthorblockA{\IEEEauthorrefmark{2}Depart. of Systems and Computer Eng., Carleton University, Ottawa, ON, Canada}\\
\IEEEauthorblockA{\IEEEauthorrefmark{3}State Key Laboratory of Networking and Switching Technology, Beijing University of Posts and Telecommunications, Beijing, P.~R.~China}\\
\IEEEauthorblockA{\IEEEauthorrefmark{4}National Mobile Communications Research Laboratory, Southeast University, Nanjing, P.~R.~China}\\
}

\maketitle

\begin{abstract}
Wireless network virtualization and information-centric networking (ICN) are two promising technologies for next generation wireless networks. Although some excellent works have focused on these two technologies, device-to-device (D2D) communications have not beeen investigated in information-centric virtualized cellular networks. {\color{black} Meanwhile, content caching in mobile devices has attracted great attentions due to the saved backhaul consumption or reduced transmission latency in D2D-assisted cellular networks. However, when it comes to the multi-operator scenario, the direct content sharing between different operators via D2D communications is typically infeasible. In this article, we propose a novel information-centric virtualized cellular network framework with D2D communications, enabling not only content caching  in the air, but also inter-operator content sharing between mobile devices.} Moreover, we describe the key components in the proposed framework, and present the interactions among them. In addition, we incorporate and formulate the content caching strategies in resource allocation optimization, to maximize the total utility of mobile virtual network operators (MVNOs) through caching popular contents in mobile devices. Simulations results demonstrate the effectiveness of the proposed framework and scheme with different system parameters.
\end{abstract}

\IEEEpeerreviewmaketitle

\section{Introduction}

The ever-increasing popularity of smart mobile devices and applications has given rise to the substantial growth in wireless  traffic \cite{LY15,MYL04}. Researchers in both academia and industry make great efforts to increase the network capacity by employing sophisticated techniques, e.g., enhancement of the physical layer capacity between transmitters and receivers, utilization of extra spectrum as well as improvement of  spectral efficiency \cite{Golrezaei2014Base-Station,XYJL12}. Nevertheless, in spite of the aforementioned advances, the spectrum efficiency is explicitly reaching its theoretical upper bound, giving rise to new requirements towards the wireless network architecture \cite{Xylomenos2014Survey}. Meanwhile, it can be observed that the majority of wireless traffic is owing to replicate downloads of some popular contents \cite{Liu2014Content-centric, Xylomenos2014Survey}. 

In this context, \emph{information-centric networking} (ICN) has been proposed as a promising technology for next generation wireless networks \cite{LYZ15}, to tackle this shift from the \emph{connection-centric} paradigm to a more \emph{content-centric} paradigm. By naming contents at the network layer, ICN is characterized by in-network caching and receiver-driven content-level delivery as well as multicast transmissions. With ICN, mobile users can directly fetch contents from nearby caches rather than from core networks (CNs), enabling not only the alleviation of traffic load across networks but also the enhancement of energy efficiency \cite{LYZ15}.

Another promising technology for next generation wireless networks is \emph{wireless network virtualization}, with which the wireless network infrastructure can be decoupled from the services and applications it provides \cite{LY15}. Towards this end, the network \emph{hypervisor}, a critical entity residing in the control plane \cite{Blenk2016Survey}, takes the responsibility of virtualizing underlying networks, allocating virtual resources and eventually managing virtual networks. Since multiple mobile virtual network operators (MVNOs) can dynamically share the physical substrate networks, the capital expenses (CapEx) and operation expenses (OpEx) of radio access networks (RANs) as well as CNs, can be reduced significantly \cite{LY15}. Besides, MVNOs running on top of virtual networks can provide some specific over-the-top services (e.g., VoIP) to subscribers, which facilitates the attraction of more users for mobile network operators (MNOs). 

The integration of ICN with wireless network virtualization can further facilitate the improvement of applications and services experienced by subscribers \cite{LYZ15}. In particular, \emph{information-centric wireless network virtualization} allows the sharing of not only the infrastructure, but also contents residing in different MVNOs, to enable the gain from not only virtualization but also in-network caching. 

{\color{black} Although some excellent works have focused on ICN and wireless network virtualization, \emph{device-to-device} (D2D) communications \cite{Asadi2014Survey} have not been incorporated in the information-centric wireless network virtualization architecture. With D2D communications, users in close proximity can directly communicate with each other via D2D links, instead of accessing base stations (BSs) exclusively. When it comes to the ICN paradigm, in spite of the smaller-sized storage (compared to that of BSs), the ubiquitous caching capability residing in mobile devices cannot be neglected due to their ubiquitous in-network distribution and ever-increasing storage size \cite{Liu2014Content-centric}. There have been some existing works (e.g., \cite{Gregori2016}) concerned with the information-centric D2D-assisted cellular networks, where the network performance could be considerably improved via D2D communications in terms of the saved backhaul consumption or reduced transmission latency. Yet, all of them are identically under the assumption that there exists only one network operator in the infrastructure. That is, all mobile devices are subscribed to and managed by the same operator. However, when referring to the multi-operator case, two users from different operators could not directly communicate with each other, due to policies and markets (i.e., economic factors) rather than technologies. Therefore, inter-operator content sharing via D2D communications are typically infeasible in legacy cellular networks, necessitating the introduction of virtualization into D2D-assisted information-centric cellular networks, to enable the immediate content delivery among subscribers from different MVNOs.}

In this article, we investigate information-centric virtualized cellular networks with D2D communications, and explore the potential opportunities and challenges. The distinctive contributions of this article are as follows:

\begin{itemize}
  \item We present an information-centric virtualized cellular network framework, enabling the content caching not only in the air, but also in mobile devices. In particular, by caching contents in mobile devices, virtualized contents can be shared by subscribers from different MVNOs.

  \item In the proposed framework, we describe the fundamental components (e.g., cellular network infrastructure, radio spectrum, MVNO, subscriber, and the hypervisor), and present the interactions among them.

  \item We formulate the content caching strategies as a utility maximization problem with D2D communications involved. This formulation enables us to quantify the gain through caching popular contents (e.g., videos) in mobile devices. Simulations are conducted with different system parameters to show the effectiveness of the proposed framework and scheme.

\end{itemize}

The remainder of this article is organized as follows: In Section \ref{sec:Framework}, we present the proposed information-centric virtualized cellular network framework with D2D communications. In Section \ref{sec:Components}, we describe the fundamental components of the proposed framework and then present the interactions among them. In Section \ref{sec:Optimization}, we optimize the resource allocation to maximize the total utility of all MVNOs with caching strategies involved in the proposed framework. In Section \ref{sec:Simulation}, we evaluate the proposed framework and algorithm through simulations. We conclude this work in Section \ref{sec:Conclusion} with future work.

\section{Information-Centric Virtualized Cellular Network Framework with D2D Communications} \label{sec:Framework}
In this section, we first introduce a general information-centric virtualized cellular network framework, then incorporate D2D communications into it.

\subsection{Information-Centric Cellular Network Virtualization}
Conventional cellular networks face the following challenges when the content retrieval technology is involved. First, it may incur a heavy outbound traffic, when contents are scattered among MNOs. That is to say, due to the proprietary nature of substrate networks, the direct content exchange between different MNOs is impractical, and thus the traffic must be transmitted in a duplicative manner. Second, for the implementation of ICN technology, the traditional cellular network framework \cite{YL01} may have to undergo a large evolution in the content format and communication protocol. However, current devices and protocols do not adapt to the new ICN paradigm.

Fortunately, \emph{network function virtualization} (NFV) \cite{LY15}, which enables the sharing of not only infrastructure but also contents among Internet service providers (ISPs), is an effective approach to implement the ICN paradigm within current cellular networks. In the traditional content delivery networking (CDN) framework, a subscriber may have to access a remote BS or even acquire the content from CNs via the backhaul, provided that nearby BSs are not operated by the same MNO with this subscriber. By contrast, with NFV, a subscriber can fetch its required content immediately from the nearby BS (which can be virtualized and shared by multiple MVNOs). As such, redundant traffic can be significantly decreased, contributing to the reduction of CapEx and OpEx of RANs as well as CNs, and the improvement of quality of experience (QoE) experienced by users. In addition, from the perspective of infrastructure providers (InPs), this also helps produce more benefits by leasing isolated virtual slices to MVNOs.

\begin{figure}[!t]
	\centering
	\includegraphics[width=4in]{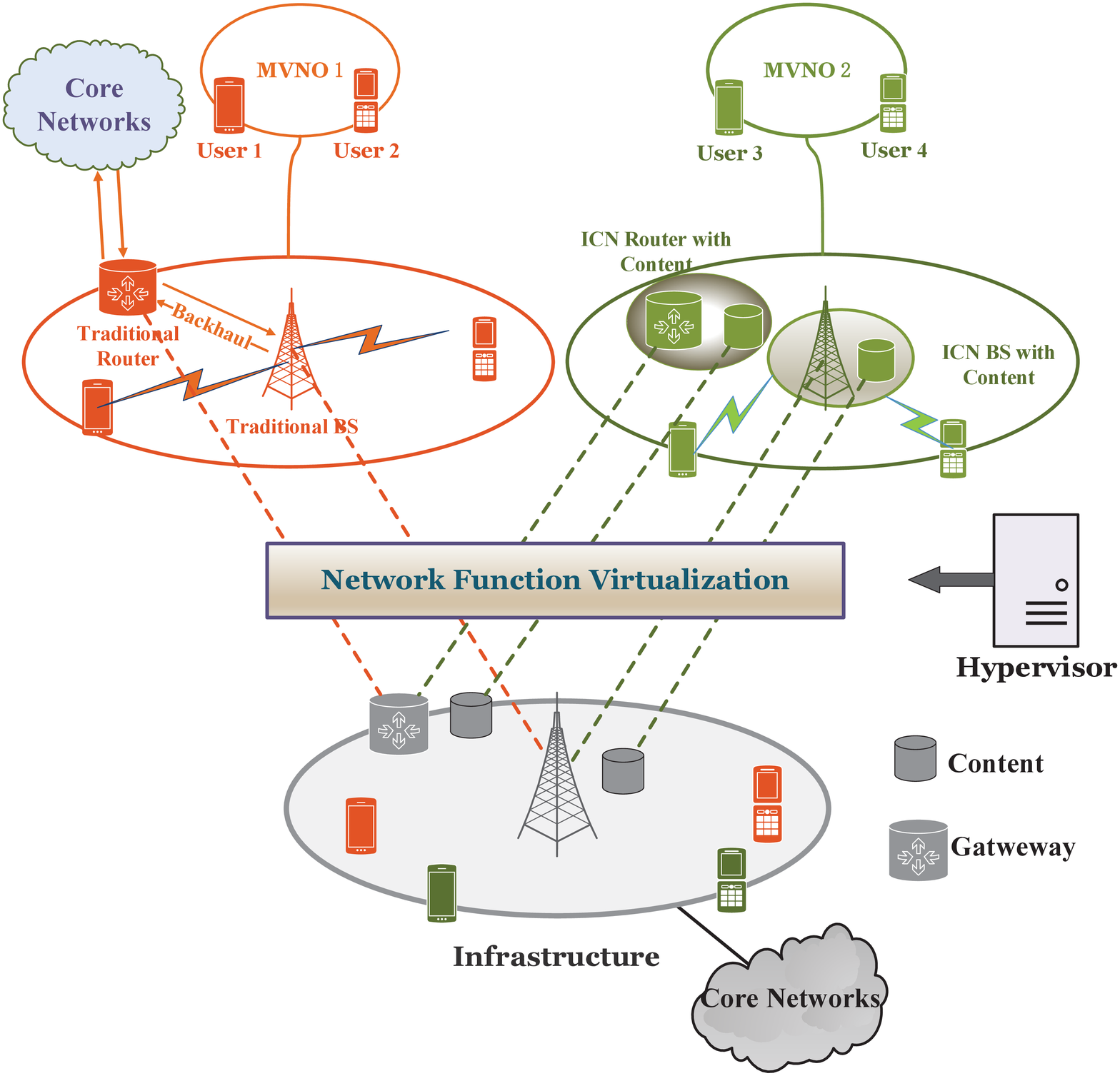}
	\caption{A general information-centric cellular network virtualization model.}
	\label{fig:Virtualization}
\end{figure}

Fig. \ref{fig:Virtualization} illustrates a general information-centric cellular network virtualization model, where the infrastructure (i.e., the substrate network) is abstracted and virtualized into two \emph{customized} virtual slices, leased to and supervised by MVNO1 and MVNO2, respectively. As shown in Fig. \ref{fig:Virtualization}, MVNO 1 runs a traditional cellular network, while MVNO 2 runs an ICN-based one. User 3 and user 4 (from MVNO 2) can directly get contents from the caching BS or router, whereas user 1 and user 2 (from MVNO 1) have to fetch contents from CNs via the backhaul. It should be noted that, although they subscribe to different MVNOs, all users can attach to the same BS owing to the virtualization of it to two \emph{virtual BSs}.

\subsection{Information-Centric Cellular Network Virtualization with D2D Communications}
The in-network storage for \emph{content caching} is an integral part of information-centric cellular networks. Most existing works assume that the caching function only runs in the BS, since the content retrieval merely arises in the downlink (i.e., BS $\rightarrow$ users) and not vice versa. Nevertheless, it is inadvisable to neglect the caching capability in mobiles devices when the D2D communication is involved, due to its massive number and ever-increasing storage size \cite{Golrezaei2014Base-Station}. Therefore, caching functions in mobile devices are non-negligible even though users are in general with the smaller caching memory compared to BSs. Moreover, the exponential growth in wireless data traffic also necessitates the ubiquitous caching in devices to offload traffic in the air.

To incorporate D2D communications into the general information-centric cellular network virtualization framework, one challenge arises: how to perform the \emph{content-level} slicing \cite{LYZ15} in D2D pairs? In general, each content residing in BSs is virtualized to several virtual slices, and thus can be shared and accessed dynamically by multiple MVNOs. And it can be physically interpreted as broadcasting the popular content to customers from different MVNOs within the same BS coverage, and embedding several virtual content slices into one physical content. However, when it comes to the content-level slicing in D2D communications, two main technical problems needed to be tackled are as follows:

\textbf{How to connect two arbitrary customers from different MVNOs?}: On one hand, in traditional D2D networks without virtualization, subscribers from different MNOs cannot directly communicate with each other even though they are in close proximity, given the absence of protocol or agreement between MNOs. On the other hand, in the aforementioned general virtualization framework, MVNOs need to pay InPs to lease the infrastructure. However, unlike the typical virtualization scenario where each BS in the infrastructure can be virtualized as {virtual nodes} for content broadcasting, the virtualization of D2D users is not straightforward since mobile devices are not {part of the infrastructure} in reality. Thus, in this article, we propose a strategy that MVNO 1 needs to pay MVNO 2 if
MVNO 1 intends to establish an immediate D2D connection between its subscriber and MVNO 2's subscriber. That is to say, one MVNO transfers  the fee (which should have been charged by InPs) from InPs to another MVNO. In this manner, the protocol is set up to facilitate the direct communication between MVNOs. Specifically, a user considered as the {D2D transmitter} can be paid by one MVNO (supposing {one-to-one D2D unicast}) or multiple MVNOs (supposing {one-to-more D2D broadcast or multicast} \cite{Cai2015Software}), depending on the D2D technology underlaying cellular networks.

\textbf{What to cache in mobile devices?}: Caching policies, determining what to cache and how to refresh the memory, are of vital importance for overall caching gains. As studied in \cite{Liu2014Content-centric}, caching works towards a trade-off between the traffic offload from BSs and the memory space in devices, namely, deciding what to cache in the limited memory space to achieve the maximum gain. Compared to the BS with the large-sized memory to cache abundant contents, mobile devices are typically equipped with small-sized storage. Thus, what to cache should be optimized in mobile devices to obtain the maximum gain in terms of the saved backhaul consumption or reduced transmission delay \cite{LYZ15}. Intuitively, users may have to cache what has the highest revisit rate for neighboring ones to associate with. Due to the dynamics of the content popularity profile and network conditions, a dynamic caching strategy is obligatory, and the hypervisor has to learn the latest popularity profile and then notifies mobile devices of what to refresh in the best way possible. However, since the popularity profile is not available in advance, it is necessary for the hypervisor to estimate the caching gain in the next interval on the basis of history demand and memory size \cite{LYZ15}. Through the dynamic refreshment at {regular time intervals}, the hypervisor is capable of keeping track of the profile variation and caching those with the highest revisit rate.

\begin{figure}[!t]
	\centering
	\includegraphics[width=5in]{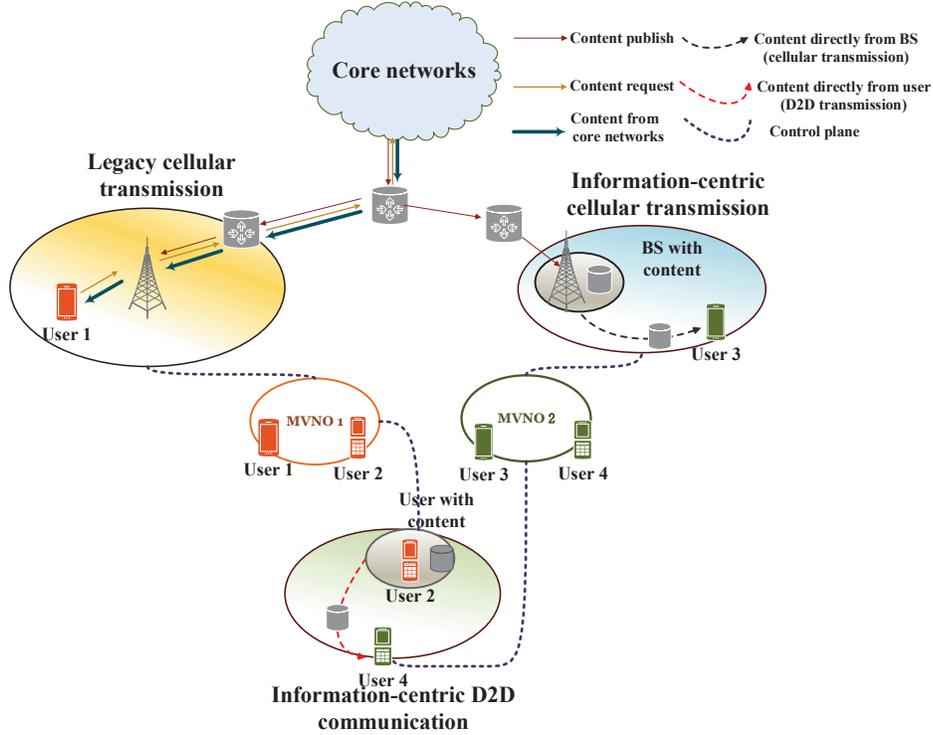}
	\caption{Information-centric cellular network virtualization with D2D Communications.}
	\label{fig:ModelD2D}
\end{figure}

In Fig. \ref{fig:ModelD2D}, we show the information-centric cellular network virtualization framework with D2D communications, where there are three concurrent content deliveries in the system. Among them, user 1 subscribed to MVNO 1 can only retrieve the content from CNs, since its required content is cached in neither BS or nearby users. User 3 from MVNO 2 can acquire its required content from the local cacheable BS, while user 4 can be immediately satisfied via the D2D link between user 2 and 4 without fetching the content from local BSs or CNs. It should be noted that even though user 2 and 4 are from different MVNOs, the direct delivery in the proposed framework can be established as long as MVNO 1 and MVNO 2 reach an agreement for content exchange. Along this line, user 2 as well as its cached content is virtualized into the virtual transmitter and virtual content, respectively, enabling the content-level slicing.

\section{Key Components of the Proposed Framework} \label{sec:Components}

\begin{figure}[!t]
	\centering
	\includegraphics[width=5in]{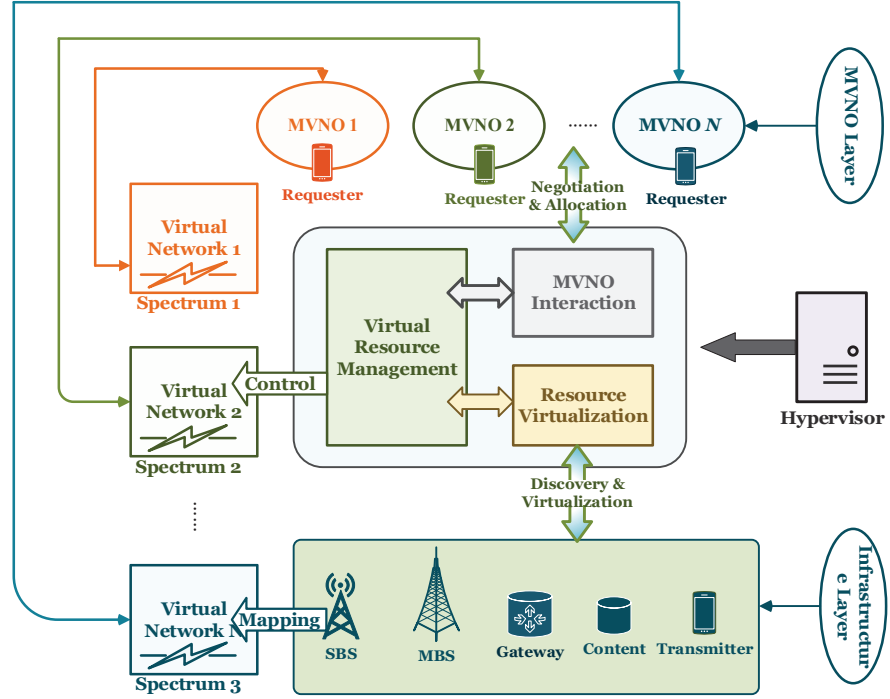}
	\caption{Key components of the proposed framework and the interactions among them.}
	\label{fig:Architecture}
\end{figure}

In this section, we describe the key components of the proposed framework, including cellular network infrastructure, radio spectrum, MVNO, subscriber, and hypervisor. The architecture of of these components is shown in Fig. \ref{fig:Architecture}, where the interactions among them are illustrated as well. In particular, as shown in Fig. \ref{fig:Architecture}, the hypervisor is further split into three \emph{function entities}, i.e., \emph{resource virtualization}, \emph{virtual network management} and \emph{MVNO interaction}. Next, each component in the framework will be described in detail.

\subsection{Cellular Network Infrastructure}

The cellular network infrastructure is the foundation of the virtualization framework and the cornerstone of virtualized slices. In conventional cellular networks, within one area, the infrastructure is typically invested by one MNO and thus can only be operated by it exhaustively. By contrast, in the proposed framework, the cellular network infrastructure is constructed and funded by InPs, which take the responsibility of managing and maintaining it as well. Therefore, InPs play a significant role in terms of the routine operation of \emph{infrastructure-as-a-service} (IaaS) \cite{LY15}.

In particular, although originating from ISPs, the content residing in cellular networks should also be considered as one element in the infrastructure, since it can be virtualized to facilitate the direct exchange between MVNOs. As a consequence, D2D transmitters with requested contents in memory should be regarded as functioning identically as those typical infrastructure elements, such as macro base stations (MBSs), small base stations (SBSs), backhaul, CNs, gateways, with the only difference lying that InPs need to forward the rent payments by MVNOs to those D2D transmitters.

\subsection{Radio Spectrum}

Two representative radio spectrum utilization cases are as follows:

\emph{\textbf{InP Ownership ---}} In this case, radio spectrum is part of InPs. After abstraction, virtualization and slicing of the cellular network infrastructure, the network hypervisor eventually assembles virtual slices with spectrum segments to realize a \emph{spectrum-level isolation} \cite{LY15} between virtual networks. It should be noted that a flexible spectrum sharing scheme among MVNOs is realized, as specified by
the negotiation between MVNOs and InPs.

\emph{\textbf{MVNO Ownership ---}} MVNO ownership indicates that radio spectrum is owned by MVNOs, namely, each operator occupies its proprietary licensed spectrum segment. Thus, each virtualized slice may have to be combined with its relevant spectrum segment to constitute a entire virtual network, customized by each operator. In this case, the spectrum sharing scheme among MVNOs is impractical unless a protocol or agreement is reached.

\subsection{MVNO}
An MVNO typically exposes two main functions. On one hand, each MVNO has to charge its subscribers for service access and content retrieval, to enable the regular operation of cellular networks. On the other hand, each MVNO also has to interface and interact with the hypervisor to negotiate the leasing price for the infrastructure. Upon being aware of the customized virtual network from the hypervisor, each MVNO can further make decisions on how to re-distribute available virtual resources to its subscribers. It is straightforward that the subscribers paying more to MVNOs tend to experience the improved QoE, while those paying less are very likely to download contents (e.g., videos and games) with a low data rate.

\subsection{Subscriber}
Subscribers are divided into two categories, i.e., {transmitters} and {requesters}, at each virtual resource allocation time interval. Transmitters are responsible for content delivery and typically regarded as substrate resources by the hypervisor, while requesters need to interact with the subscribed MVNOs to set the subscription price and hence function as {routine subscribers}.

Each requester can decide whether or not to cache the acquired content, whereas each transmitter remains the content in its memory at the current interval. Yet, over the long run, even a transmitter cannot refresh the caching memory at time interval $t$, it is likely that it will convert to a requester at interval $t + 1, t + 2, \cdots$. Therefore, the role of subscribers is not predetermined and each subscriber is capable of transitioning between the two modes at a large time-scale.

\subsection{Hypervisor}
The hypervisor, supported by the host operating system \cite{Blenk2016Survey}, monitors the virtual networks running on top of the infrastructure, and allocates substrate network resources to each customized virtual slice. There are three function entities residing in the hypervisor, namely, \emph{resource virtualization}, \emph{MVNO interaction} and \emph{virtual network management}, which are described below.

\emph{\textbf{Resource Virtualization ---}} The resource virtualization entity interfaces with InPs, and takes the responsibility of discovering and abstracting substrate resources into {virtual elements}. Then, it packages and dispatches the information of all available virtual elements to the unit virtual network management for the slicing and assembling of customized virtual networks. Meanwhile, it is also responsible for embedding virtual slices into substrate networks, notifying substrate nodes of the resource allocation policies (e.g., the transmit power per node and the user set that each node serves).

\emph{\textbf{MVNO Interaction ---}} To enable the information exchange between the hypervisor and MVNOs, a unit MVNO interaction is deployed as a guest operating system, to negotiate about the leasing price. Meanwhile, the quality of service (QoS) requirements as well as content requests of all subscribers are forwarded by MVNOs to this entity. Then, it delivers the collected data to the unit virtual network management to enable the proprietary virtual network customized by each MVNO. In addition, it also takes the responsibility of disseminating allocation results to each subscriber (e.g., the associated BS that each user attaches to).

\emph{\textbf{Virtual Network Management ---}} Leveraging the information from both resource virtualization and MVNO interaction, the entity virtual network management can flexibly slice and aggregate the observed virtual elements into multiple integral virtual networks, as specified by MVNOs. Meanwhile, the other two entities receive and process the allocation-related signaling from the virtual network management, and then forward it to MVNOs and InPs, respectively.

\section{Resource Allocation in the Proposed Framework} \label{sec:Optimization}
The hypervisor not only customizes virtual networks as specified by network operators, but also embeds those virtualized slices into substrate networks, namely, indicating the user set that each transmitter serves, the spectrum segment that each link occupies, the cache refreshment policy, etc. In essence, the proposed framework presents two unique techniques, namely, \emph{D2D communications between different operators} and \emph{caching capability in requesters}, that can be leveraged in algorithm development.

\emph{\textbf{D2D Communications between Different Operators ---}} Consider the single-hop and one-to-one (i.e., unicast) D2D communications in cellular networks. As mentioned above, D2D links can operate between subscribers from different MVNOs, thus enabling sharing and virtualization of contents. Therefore, we should distinguish D2D links from traditional ones, and then formulate the utility function corresponding to each potential link $j \rightarrow (m,i)$, where $j \in \mathcal{J} = \{1, \ldots, J\}$ denotes the D2D transmitter index, $j = 0$ denotes the cellular transmission, and $(m,i)$ is the $i$-th subscriber of MVNO $m$, respectively. Here, if $i$ and $j \in \mathcal{J}$ are from different MVNOs, then content sharing is allowed. Moreover, we introduce the spectrum allocation variable $y_{j}^{mi}$ to indicate the the spectrum fraction allocated to this link by the hypervisor, and $r_j^{mi}$ is the net gain involving both the revenue from subscribers and the leasing cost charged by InPs assuming that the entire bandwidth is assigned to this link \cite{LYZ15}.

\emph{\textbf{Caching Capability in Requesters ---}} With respect to D2D links, caching refreshment in receivers must be incorporated in the algorithm design to facilitate the content retrieval for neighboring users. Denote the content requested by $(m,i)$ as $c_{mi}$. We further introduce the binary variable $z_{j}^{mi}$ to denote whether or not the content sent by $j$ is cached by requester $(m, i)$. Meanwhile, $v^{mi}$ is utilized to indicate whether or not BS caches content $c_{mi}$. In addition, due to requesters' limited memory space, the cache refreshment has to be executed continuously to track the popularity of cached contents.

The resource allocation scheme by the hypervisor should exploit these two natures to benefit from not only D2D communications, but also caching gains. Along this line, the utility function of the link between requester $(m,i)$ and transmitter $j$ should be designed as

\begin{equation}\label{Utility}
 u_j^{mi}
=  y_j^{mi} r_j^{mi}
+  z_j^{mi} (\phi_{mi} e_{c_{mi}} - \psi_{mi} s_{c_{mi}}) + v^{mi} (\phi_{mi} e_{c_{mi}} - \psi_j^{mi} s_{c_{mi}}),
\end{equation}
where $e_{c_{mi}} $ is the estimated backhaul bandwidth reduction via caching, and $s_{c_{mi}}$ denotes the storage space necessary for $c_{mi}$. Moreover, $\phi_{mi}$ and $\psi_{mi} (\psi_j^{mi})$ are the unit avenue that MVNO $m$ charges users requesting content $c_{mi}$ and the unit price that MVNO $m$ should pay requester $(m,i)$ (transmitter $j$) for the content storage, respectively. It should be noted that the third term relates to traditional cellular communications while the second term represents the caching gains via D2D transmissions.

All available resources, including transmitter, spectrum and caching memory, should be optimized such that the total utility of all MVNOs seen by the hypervisor is maximized, which can be mathematically formulated as
\begin{equation} \label{sum_utility}
\max_{y_j^{mi}, z_{j}^{mi}, v^{mi}} \sum_{m \in \mathcal{M}} \sum_{i \in \mathcal{I}_m} \sum_{j \in \mathcal{J}_0} u^{mi}_j,
\end{equation}
where $\mathcal{J}_0 = \mathcal{J} \cup \{0\}$. It should be noted that Eq. \ref{sum_utility} is in essence equivalent to maximizing 

\noindent $\sum_{j \in \mathcal{J}_0} \sum_{m \in \mathcal{M}} \sum_{i \in \mathcal{I}_m} u^{mi}_j$, which facilitates the distribution of computing burden on the hypervisor to each transmitter $j$ to reduce the computational complexity, based on the recent advances on \emph{alternating direction method of multipliers} (ADMM) \cite{Boyd2011Distributed}. As such, the associated subproblem is separately solved by each transmitter and then updated to the hypervisor for collaboration.

\section{Performance Evaluation} \label{sec:Simulation}
In this section, we highlight the performance of proposed ADMM-based distributed algorithm (denoted as distributed algorithm for simplicity) with caching and D2D in terms of total utility of MVNOs, by comparing it with three other algorithms listed as follows:

\begin{itemize}
  \item Centralized algorithm with (w.) caching and D2D (denoted as optimal strategy), where the hypervisor collects channel state information (CSI) and content distribution information from all transmitters and then optimizes the resource allocation.
  \item Distributed algorithm without (w.o.) caching but w. D2D (denoted as w.o. caching).
  \item Distributed algorithm w. caching but w.o. D2D (denoted as w.o. D2D).
\end{itemize}

{\color{black} We conduct simulations within one cell (operated by one InP), with a radius of 500 m and the BS located at coordinates (0, 0). 65 subscribers (30 transmitters and 35 routine subscribers) are uniformly distributed within the coverage and 4 MVNOs run on top of the infrastructure, i.e., each subscriber accesses to any MVNO with a probability of 25\%. Meanwhile, the overall bandwidth is divided into two orthogonal parts: BS DL ($B_{dl} = 10$ MHz) for cellular communications and BS UL ($B_{ul} = 5$ MHz) for D2D communications. As such, there exists no co-channel interference of D2D communications on cellular transmissions. Moreover, the Zipf distribution $q_c = \frac{C}{c^\epsilon}, C = (\sum_c 1 / c^\epsilon)^{-1}$ with $\epsilon = 1.5$ \cite{Shanmugam2013FemtoCaching} is leveraged to model the popularity profile of contents, and 5 kinds of contents (with equal size) are distributed in the simulated network. The BS's storage size reaches 5 (contents), while the mobile device's storage size is limited to 1. In particular, in the initial content placement, it is assumed that each content is held by the BS with a probability of 50\%, while each D2D transmitter has a probability of 20\% to hold any of 5 contents. In addition, the transmit powers of BS and D2D transmitter are 46dBm and 24dBm, respectively. The path loss model $35.3 + 37.6 \log (d(m))$ is designated. Besides, we consider the log-normal distribution with standard deviation of 8dB for the shadowing, and exponential distribution with unit mean for the fast fading, as specified in \cite{Feng2013Device}.}

\begin{figure}[!t]
	\centering
	\includegraphics[width=4.in]{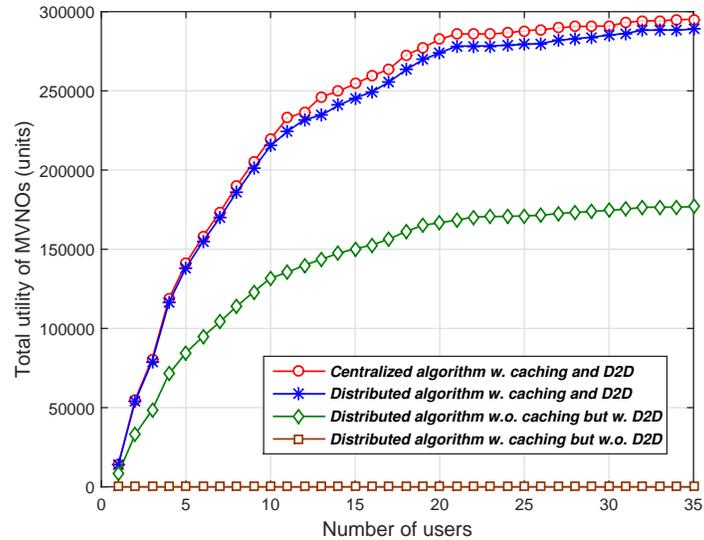}
	\caption{The total utility of MVNOs. (There are 30 D2D transmitters, the average data rate requirement per user is 2 Mbps, the average size per content is 2 Mb, and there are 2 types of contents.)}
	\label{UserNum}
\end{figure}

Fig. \ref{UserNum} compares the resource allocation algorithms in terms of total utility of MVNOs. Except for the algorithm w.o. D2D, a monotonically increasing utility is observed for all three other algorithms, namely, the total utility obtained by MVNOs increases with the increase of the number of requesters. That is due to the fact that a network incorporating more receivers will introduce multi-user diversity gain. Moreover, the total utility obtained by the distributed algorithm is only 2.8\% lower than that of the optimal strategy. Note that, for the algorithm w.o. D2D, the caching function is only available in the BS due to the non-existence of D2D communications, resulting in the least utility.

\begin{figure}[!t]
	\centering
	\includegraphics[width=4.in]{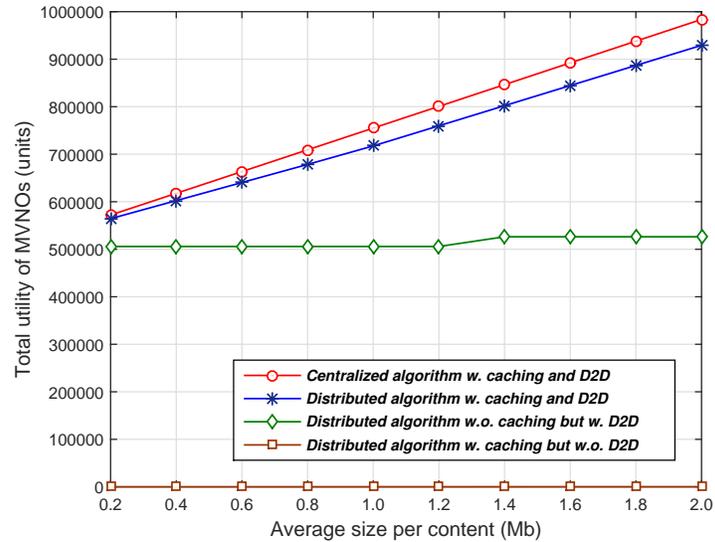}
	\caption{The total utility of MVNOs. (There are 10 D2D transmitters and 20 requesters, the average data rate requirement per user is 2 Mbps, and there are 5 types of contents.)}
	\label{ContSize}
\end{figure}

In Fig. \ref{ContSize}, we compare the performance of resource allocation algorithms with different average size of contents. The total utility of both the optimal strategy and distributed algorithm is proportional to the average size. The optimal strategy has the highest total utility (providing an upper-bound utility, which is approximately 5.3\% higher than the distributed one.), while the algorithm w.o. D2D has the lowest one. Meanwhile, the utility of both the scheme w.o. caching and w.o. D2D almost remains constant as the average size increases, which can be interpreted as that the majority of saved backhaul bandwidth consumption comes from caching popular contents, rather than directly fetching existing ones from BS and D2D transmitters.

\begin{figure}[!t]
	\centering
	\includegraphics[width=4.in]{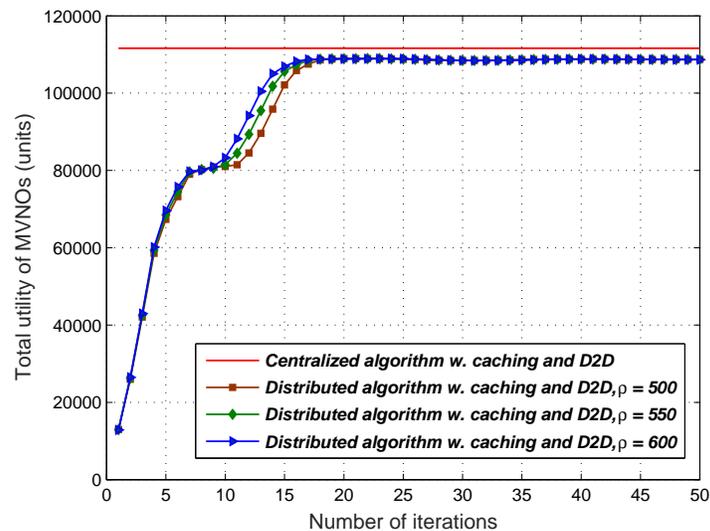}
	\caption{Convergence of the ADMM-based algorithm under different values of $\rho$.}
	\label{fig:Convergence}
\end{figure}

Fig. \ref{fig:Convergence} demonstrates the convergence of the proposed algorithm with $\rho = 500, 550,$ and $600$ ($\rho$ is the \emph{penalty parameter} for the ADMM algorithm). It can be observed that the proposed ADMM-based algorithm with different values of $\rho$ eventually converges to the same total utility. Meanwhile, all the curves converge to a stable solution monotonically within 20 iterations. The difference of $\rho$ only takes effect on the convergence speed but not on the convergence value. In particular, $\rho = 600$ is the fastest when converging to the stable solution,  while $\rho = 500$ is the slowest. Nevertheless, the difference is not significant.

\section{Conclusions and Future Work}\label{sec:Conclusion}
In this paper, we proposed to incorporate D2D communications into information-centric virtualized cellular networks. First, we developed an information-centric virtualized cellular network framework with D2D communications, considering the caching gains not only in BSs but also in ubiquitous mobile devices. Then, we described the key components in the proposed framework in detail and presented the interactions them. In addition, to quantify the caching gains, we designed a new utility function associated with D2D links, involving the revenue from user access as well as the estimated gain via caching popular contents. Simulation results showed that MVNOs can benefit from both information-centric wireless virtualization and D2D communications, and the proposed ADMM-based distributed algorithm can achieve near-optimal performance and good convergence property. Future work is in progress to extend the unicast D2D transmission to the multicast D2D case, as well as the single-cell to multi-cell scenario, in the proposed framework.

\section{acknowledgement}
This work was supported in part by the National Science Foundation under Grant 91338115, Grant 61231008, Grant 61372089, and Grant 61571108; by the National S\&T Major Project under Grant 2015ZX03002006; by the Fundamental Research Funds for the Central Universities under Grant WRYB142208 and Grant JB140117; by the Program for Changjiang Scholars and Innovative Research Team in University under Grant IRT0852; by the 111 Project under Grant B08038; by the Natural Sciences and Engineering Research Council of Canada; by the China Scholarship Council; and by the Open Foundation of State key Laboratory of Networking and Switching Technology (Beijing University of Posts and Telecommunications) under Grant SKLNST-2016-2-14. 

\bibliographystyle{IEEEtran}
\bibliography{virtualization_mag}

\clearpage
\begin{IEEEbiographynophoto}{Kan Wang} (kanwangkw@outlook.com) received the B.S. degree in broadcasting and television engineering from Zhejiang University of Media and Communications, Hangzhou, China, in 2009, and the Ph.D. degree in military communications with the State Key Lab of ISN, Xidian University, Xian, China, in 2016. From Oct. 2014 to Oct. 2015, he was also with Carleton University, Ottawa, ON, Canada, as a visiting scholar funded by China Scholarship Council (CSC). From Sept. 2016, he has been with the School of Internet of Things Engineering, Jiangnan University, Wuxi, China. His current research interests include 5G cellular networks, resource management, and massive IoT.
\end{IEEEbiographynophoto}

\begin{IEEEbiographynophoto}{F. Richard Yu} (Richard.Yu@carleton.ca) received the PhD degree in electrical engineering from the University of British Columbia (UBC) in 2003. From 2002 to 2006, he was with Ericsson (in Lund, Sweden) and a start-up in California, USA. He joined Carleton University in 2007, where he is currently an Associate Professor. He received the IEEE Outstanding Leadership Award in 2013, Carleton Research Achievement Award in 2012, the Ontario Early Researcher Award (formerly Premiers Research Excellence Award) in 2011, the Excellent Contribution Award at IEEE/IFIP TrustCom 2010, the Leadership Opportunity Fund Award from Canada Foundation of Innovation in 2009 and the Best Paper Awards at IEEE ICC 2014, Globecom 2012, IEEE/IFIP TrustCom 2009 and Int'l Conference on Networking 2005. His research interests include cross-layer/cross-system design, security, green IT and QoS provisioning in wireless-based systems.
	
He serves on the editorial boards of several journals, including Co-Editor-in-Chief for Ad Hoc $\&$ Sensor Wireless Networks, Lead Series Editor for IEEE Transactions on Vehicular Technology, IEEE Communications Surveys $\&$ Tutorials, EURASIP Journal on Wireless Communications Networking, Wiley Journal on Security and Communication Networks, and International Journal of Wireless Communications and Networking. He has served as the Technical Program Committee (TPC) Co-Chair of numerous conferences. Dr. Yu is a registered Professional Engineer in the province of Ontario, Canada.
\end{IEEEbiographynophoto}

\begin{IEEEbiographynophoto}{Hongyan Li} (hyli@xidian.edu.cn) received the M.S. degree in control engineering from Xi'an Jiaotong University, Xi'an, China, in 1991 and the Ph.D. degree in signal and information processing from Xidian University, Xi'an, in 2000. She is currently a Professor with the State Key Laboratory of Integrated Service Networks, Xidian University. Her research interests include wireless networking, cognitive networks, integration of heterogeneous network, and mobile ad hoc networks.
\end{IEEEbiographynophoto}

\begin{IEEEbiographynophoto}{Zhengquan Li} (lzq722@sina.com) received the B.S. degree from Jilin University of Technology in 1998, the M.S. degree from University of Shanghai for Science and Technology in 2000, and the Ph.D. degree in Circuit and System from Shanghai Jiaotong University in 2003. He joined Jiangnan University in 2016, where he is currently a professor. He is also currently a post-doctor in the National
Mobile Communications Research Laboratory, Southeast University. His current research interests include space time coding, cooperative communications, and massive MIMO.	
\end{IEEEbiographynophoto}

\end{document}